\documentclass[12pt,preprint]{aastex}

\slugcomment{Draft:  \today}

\lefthead{Rodgers et~al.}
\righthead{Corrections to the $u'g'r'i'z'$ Filter System}

\begin{document}

\title{Improved $u'g'r'i'z'$ to $UBVR_CI_C$ Transformation Equations for 
Main Sequence Stars}

\author{
Christopher T. Rodgers\altaffilmark{1},
Ron Canterna\altaffilmark{1},
J. Allyn Smith\altaffilmark{1,2},
Michael J. Pierce\altaffilmark{1}, and
Douglas L. Tucker\altaffilmark{3}
}

\altaffiltext{1}{University of Wyoming,
            Department of Physics \& Astronomy,
            Laramie, WY 82071
  \label{Wyoming}}
\altaffiltext{2}{Los Alamos National Laboratory, ISR-4, D448,
            Los Alamos, NM 87545}
\altaffiltext{3}{Fermi National Accelerator Laboratory,
                 P.O. Box 500, Batavia, IL 60510
  \label{Fermilab}}

\begin{abstract}
We report improved transformation equations between the
$u'g'r'i'z'$ and $UBVR_CI_C$ photometric systems.  Although the details
of the transformations depend on luminosity class, we find a typical
rms scatter on the order of 0.001 magnitude if the sample is limited to 
main sequence stars.  Furthermore, we find an accurate transformation requires
complex, multi-color dependencies for the bluer bandpasses.  
Results for giant stars will be reported in a subsequent paper.

\end{abstract}


\section{Introduction}
Stellar photometry has a long and successful history in the study of
stellar properties.  However, high precision photometry is
necessary in order to enable accurate measurements of extinction,
metallicity, and tests of stellar evolutionary models.  Several filter systems 
have been developed over the last few decades with strengths and weaknesses 
associated with each (see \citep{Straizys99} for a review).  Recently, the
Sloan Digital Sky Survey (hereafter SDSS; \citep{York00}) has greatly
enhanced the available photometric data for Galactic stars and thereby
promises to significantly add to our understanding of the Galaxy's stellar
populations and formation history.

Although the $u'g'r'i'z'$ filter system was defined by 
\citet[][F96]{Fukugita96} for the study of large scale structure and quasars 
in the northern galactic cap, these data have also provided a wealth of
photometric data for Galactic stars \citep[e.g.][]{Newberg99}.  Given the great 
potential of the SDSS data, we have begun a photometric study of star clusters
in the $u'g'r'i'z'$ filter system \citep{Rider04,Moore05,Smith04}.  
However, most of the existing photometric data and stellar
evolutionary models are limited to the Johnson-
Cousins $UBVR_CI_C$ filter system.  Thus accurate transformation equations are
necessary in order to evaluate the utility of the $u'g'r'i'z'$ filter system.

As part of the development of the SDSS photometric system, F96 derived 
preliminary transformation equations between the $UBVR_CI_C$ and 
$u'g'r'i'z'$ filters.  These transformations were derived from the expected
bandpasses, the predicted response of the SDSS detectors, and the spectral 
energy distribution of a metal poor, F subdwarf, specifically 
BD+17$^{\circ}$4708.  This star was chosen as the primary standard due to the 
availability of high precision, absolute spectrophotometry (see F96).  The 
initial transformation equations of F96 were primarly intended to enable the
simulation of the colors of various celestial objects.  As a result, F96 
assumed that the color transformation equations (e.g. $u' - g'$ vs. \ub) 
would be linear between the two photometric systems.  

More accurate empirical transformations were derived by \citet{Smith02}
from $u'g'r'i'z'$  photometry of a subset of the $UBVR_CI_C$ standard stars 
from \citet{Landolt92}.  \citet{Smith02} also assumed a linear transformation 
and made no distinction between luminosity classes among their standard stars.
Finally, more complex transformations for $UBV$ to $u'g'r'$ filters were presented by 
\citet{Karaali05}.  These equations include two Johnson colors for the $u' - g'$
and $g' - r'$ relations, but do not make a distinction between luminosity classes.
In this paper we re-examine \citet{Smith02} data with the goal of deriving more accurate
transformation equations between these two photometric systems.  In \S2 we
describe our analysis procedure and \S3 contains our results.  
In \S4 we present our
conclusions and describe our ongoing efforts to expand the analysis to
other luminosity classes. 

\section{Transformation Equations Between the $u'g'r'i'z'$ and $UBVR_CI_C$ 
Systems}

\subsection{Astrophysical Issues}

\begin{figure}
\includegraphics[scale=1.0]{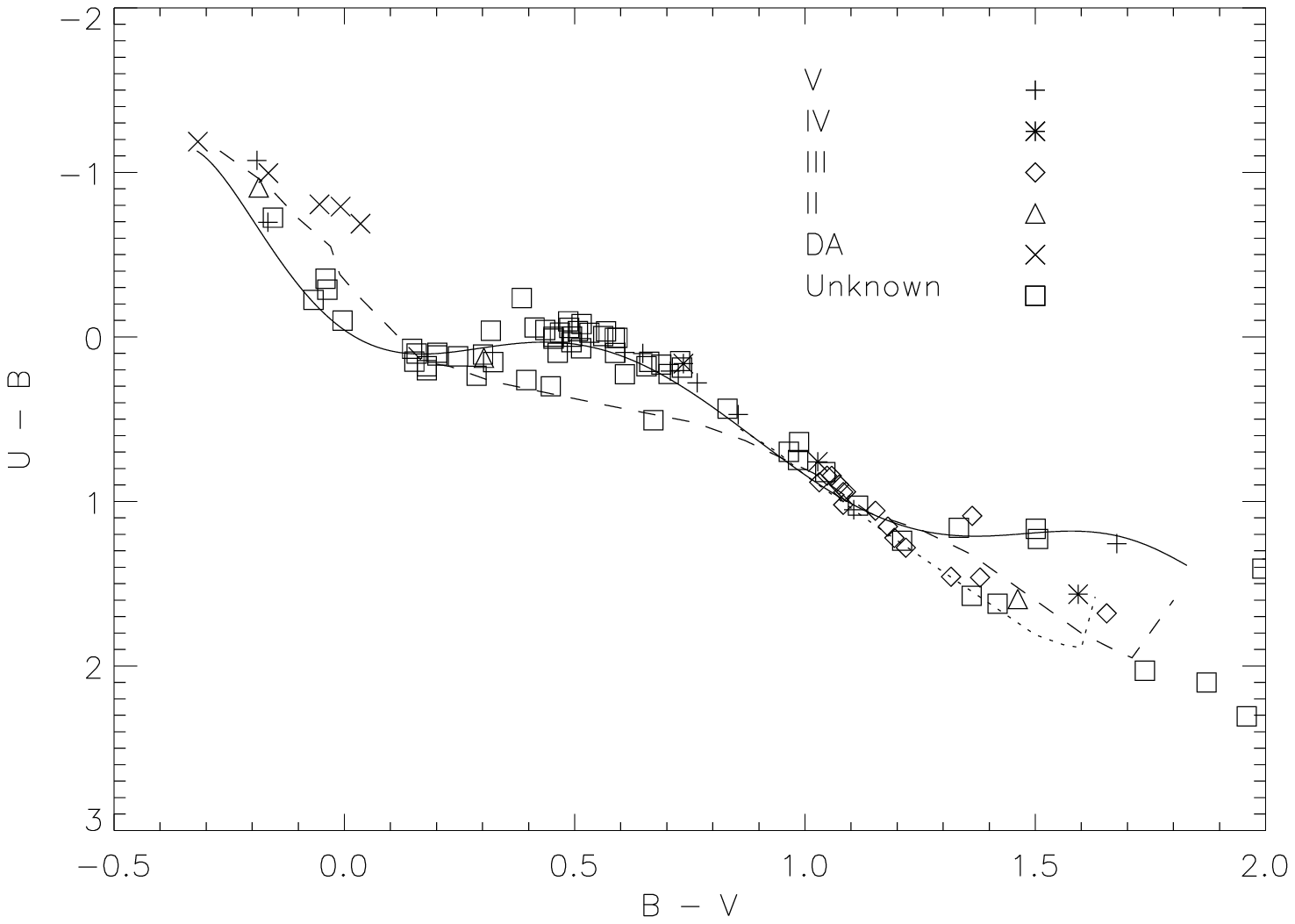}
\caption{A $UBV$ color-color diagram of the data set used in the 
\citet{Smith02} transformation equations.  The solid line is the 
smoothed ZAMS relation (luminosity class V), the dotted line is the giants 
relation
(luminosity class III), and the dashed line is the supergiants relation
(luminosity class I) taken from \citet{Johnson66}.  The different symbols
give the corresponding known and unknown luminosity classes for the
stars from \citet{Smith02}.  Using the luminosity class
V stars and well placed but unknown spectral class stars which lie along 
the ZAMS, 
we will
create new transformation equations that satisfy these data.  Note that 
Ru-152 (-0.190, -1.073) will not be included due to high reddening.
\label{smithdata}}
\end{figure}

The data (Table 1) used for this analysis is drawn from the available photometry
of the equatorial region
standard stars in common between the $UBVR_CI_C$ \citep{Landolt92} 
and the $u'g'i'r'z'$ filter systems \citep{Smith02}.  Figure~\ref{smithdata}
show these data along with the color-color relations for stars of various
luminosity classes \citep{Johnson66}.  The latter were slightly smoothed
and transformed onto the Cousins system \citep{Cousins76} using 
the transformations
derived by \citet{Fernie83}.  Figure~\ref{smithdata} illustrates the 
dependence of the transformations upon luminosity class.  Due to the limited
number of known luminosity class stars, we assumed that stars with unknown
luminosity class that lie along the ZAMS are of luminosity class V.  The 
stars with unknown luminosity class that lie in the regions of luminosity
degeneracy (where the luminosity class relations cross each other) are
still considered unknown.  The star with one exception to all of these restrictions
is Ru-152.  This star is published as an O5V, but since it lies so close to
the supergiant relation, it must either be evolved or it is highly reddened.
In any case, this star must be considered unknown as well.  Since the number  
of available supergiants and giants is limited, this paper will only address 
the transformation for main sequence (luminosity class V) stars.  

\placefigure{smithdata}
\placetable{stardata}

\begin{figure}
\includegraphics[scale=0.9]{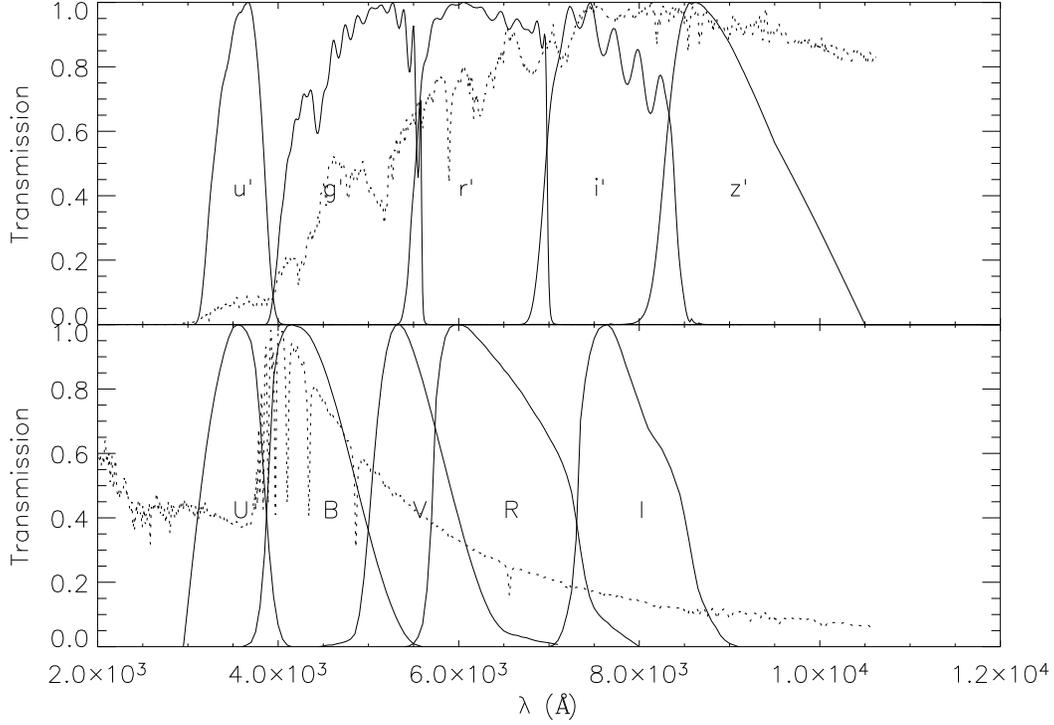}
\caption{The normalized transmission curves for $UBVR_CI_C$ and
$u'g'r'i'z'$ filter systems.  Overplotted in the upper and lower panels is a M0 V and
A0 V spectral energy distributions.  Some problems that will arise from
the physical aspects of these stars occur from first the Balmer discontinuity, second
the TiO bands in latter type stars (i.e. M0 V), and third the change of the main opacity source
at $T_eff = 10,000K$.  The Balmer discontinuity will effect the $u' - g'$ and $U - B$ colors.
The TiO bands will effect $r' - i'$, $r' - z'$, $V - R_C$, and $R_C - I_C$ colors.  The change
in opacity source at $T_eff = 10,000K$ will effect the $u' - g'$, $U - B$, and $g' - r'$ colors.
\label{filters}}
\end{figure}

Inspection of the normalized response filter curves shown in 
Figure~\ref{filters} suggests that
the system transformation relations should be a function of two colors
for most of the filter combinations.  That is,  
the $u'$ filter is closely related to the $U$ filter, but the $g'$
filter bandpass spans both $B$ and $V$ filters.  Thus,
the $u' - g'$ color should be dependent on not only the $U - B$ color, but also
the $B - V$ color.  Likewise, because the $V$ filter samples portions of 
both $g'$ 
and $r'$ filters, the $g' - r'$ color should depend on both
the $B - V$ and $V - R_C$ colors. 
The situation for the $i'$ and $z'$ filters and associated colors is less
clear since the $z'$ bandpass does not lie within the Johnson system and
is a function of the detector being used and will vary from instrument 
to instrument.

It is also evident from Figure~\ref{filters} colors involving the redder
bandpasses of both filter systems will be affected by  
the TiO molecular bands present in later spectral types (cooler stars).  
Unfortunately, there is a deficiency in the available data
for stars of later spectral type (see Figure~\ref{smithdata}).  Therefore,
our analysis will be further restricted to stars with $B - V < 1.3$ 
(see Table~\ref{stardata}).

\placefigure{filters}

\subsection{Analysis Procedure}

The transformation equations for each color were obtained using the {\tt IDL} 
{\tt REGRESS} (two dimensional, multi-variable fitting) and {\tt POLY$\_$FIT} 
(two dimensional fitting) programs.
Each $u'g'r'i'z'$ color was fit using both 
single and multi-color combinations of Johnson colors. Although higher
order fits were examined, the first order fits were found to be
suffecient. The chi-square is examined 
for each fit and the relative error in each coefficient in order to minimize
the statistical errors.
Finally, we did a ``by eye'' inspection of the fits 
to verify that the results made physical sense.

\section{Results}

We now address our results for each $u'g'r'i'z'$ color combination individually.

\subsection{The $u' - g'$ Equation}

The $u' - g'$ color is very complex in terms of the $UBV$ filters.  
The Balmer discontinuity for MS stars affect the color in $U - B$.  When 
the opacity source changes at A0 type stars, the $U - B$ color makes a 
dramatic change in color.  This effect also occurs in $u' - g'$ since 
the filters span the Balmer Jump as well.  Moreover, the $g'$ filter 
is sensitive to temperature changes in the star.  $B - V$ is also
sensitive to temperature, so the $u' - g'$ color is temperature dependent.  
With these two factors in mind, we used a first order mulit-variable 
polynomial as the basis for this first transformation equation. 

Since we are using main sequence (MS) stars, we will use the Morgan \& Keenan 
ZAMS \citep{Johnson66} in all subsequent 
figures to show the validity of each transformation by color.  Our 
transformation equation for the MS stars in $u' - g'$ is:

\begin{eqnarray}
u' - g' & = & 1.101 (\pm 0.004) (U - B) + 0.358 (\pm 0.004) (B - V) + 0.971\ .
\end{eqnarray}

\begin{figure}
\includegraphics[scale=1.0]{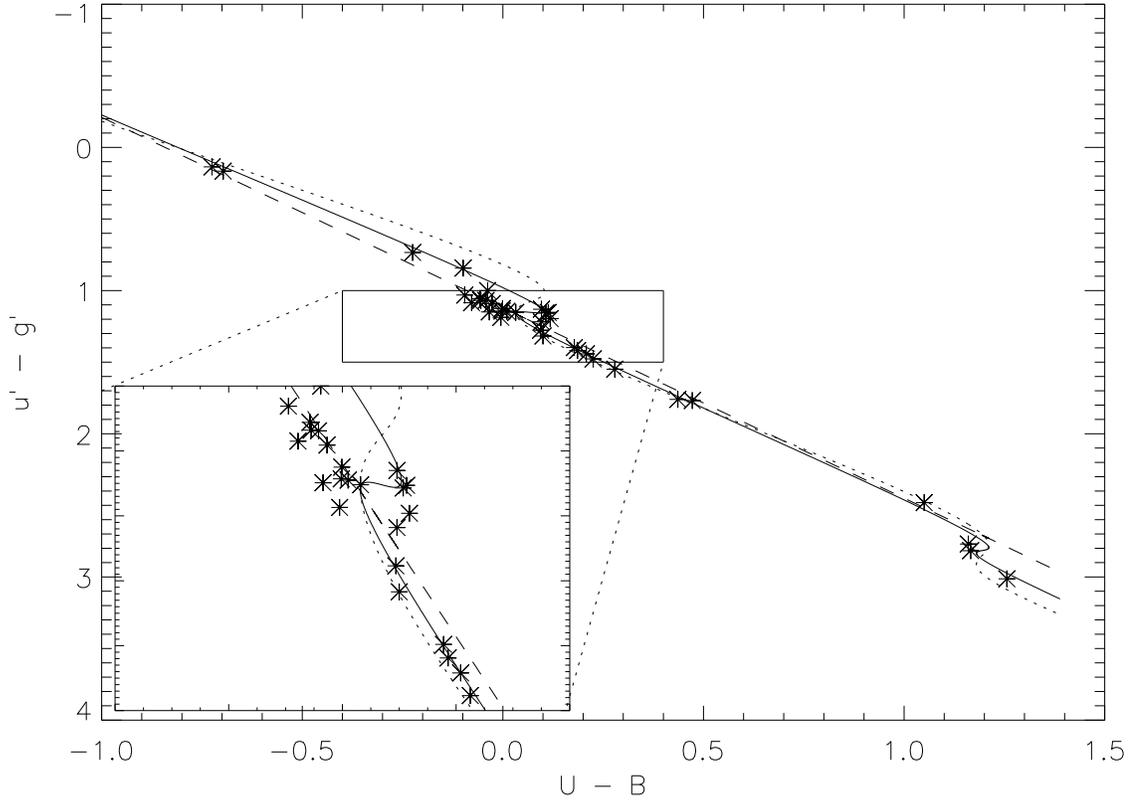}
\caption{The $u' - g'$ transformation equation fit shown here with a two 
dimensional cut of $u' - g'$ vs. \ub\ . Figure~\ref{trans_bv} gives the $u' - g'$ vs. \bv\
two dimensional cut.  The solid line is this paper's fit, the dotted
line is the \citet{Karaali05} fit, and the dashed line is the
\citet{Smith02} fit for the $u' - g'$ fit.  The inset is an expanded view of the
Balmer Jump region located in the boxed region.
Notice in the zoomed in section that the Balmer discontinuity region of
 the $u' - g'$ vs. \ub\ has been accurately fit by the new transformation equation.
\label{trans_ub}}
\end{figure}

\begin{figure}
\includegraphics[scale=1.0]{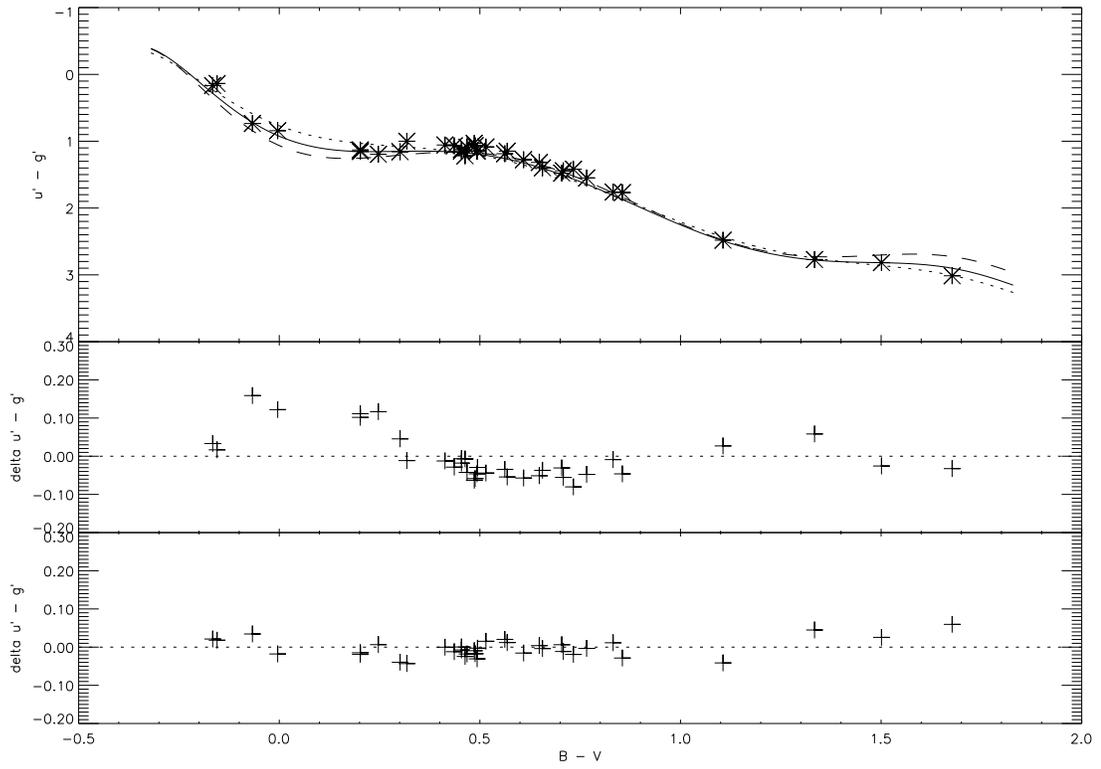}
\caption{The $u' - g'$ transformation equation fit shown here with a two 
dimensional cut of $u' - g'$ vs. \bv\ . The lines seen in the top panel are the
same as in Figure~\ref{trans_ub}.  The bottom panel gives the residuals for
both the \citet{Karaali05} fit (middle) and this paper's fit (bottom).
\label{trans_bv}}
\end{figure}

Figures~\ref{trans_ub} and \ref{trans_bv} shows the ZAMS fit in the both the $u' - g'$ 
vs. $U - B$ and $u' - g'$ vs. $B - V$ color-color planes with residuals for comparison of \citet{Karaali05}. 
As an illustration
that our new fit is more accurate, we have properly fit the Balmer 
discontinuity.  
The $u' - g'$ vs. $B - V$ plane resembles a standard color-color diagram 
with the data nicley fitting the ZAMS when the residuals are compared.  
The chi-square to each final color fit are givin in Table~\ref{chisqr}.  

\placefigure{trans_ub}
\placefigure{trans_bv}
\placetable{chisqr}

\subsection{The $g' - r'$ Equation}

The $r'$ filter maps most directly to the $R$ filter while the $g'$ 
filter overlaps both the $B$ and $V$ filters.  Therefore, we would
expect a strong dependance of $g' - r'$ on both $B - V$ and $V - R_C$.  
But before fitting this data set, the physical 
properties of extremely red stars must be taken into account.  In 
stars with $V - R_C >$ 0.8, molecular TiO bands start to dominate the 
stellar profile.  Since our data set contains only three stars this red 
an accurate transformation determination cannot be obtained after $V - R_C$ 
= 0.8.
Therefore, these 
three data points have been dropped from the final $g' - r'$ fit.  

\begin{figure}
\includegraphics[scale=1.0]{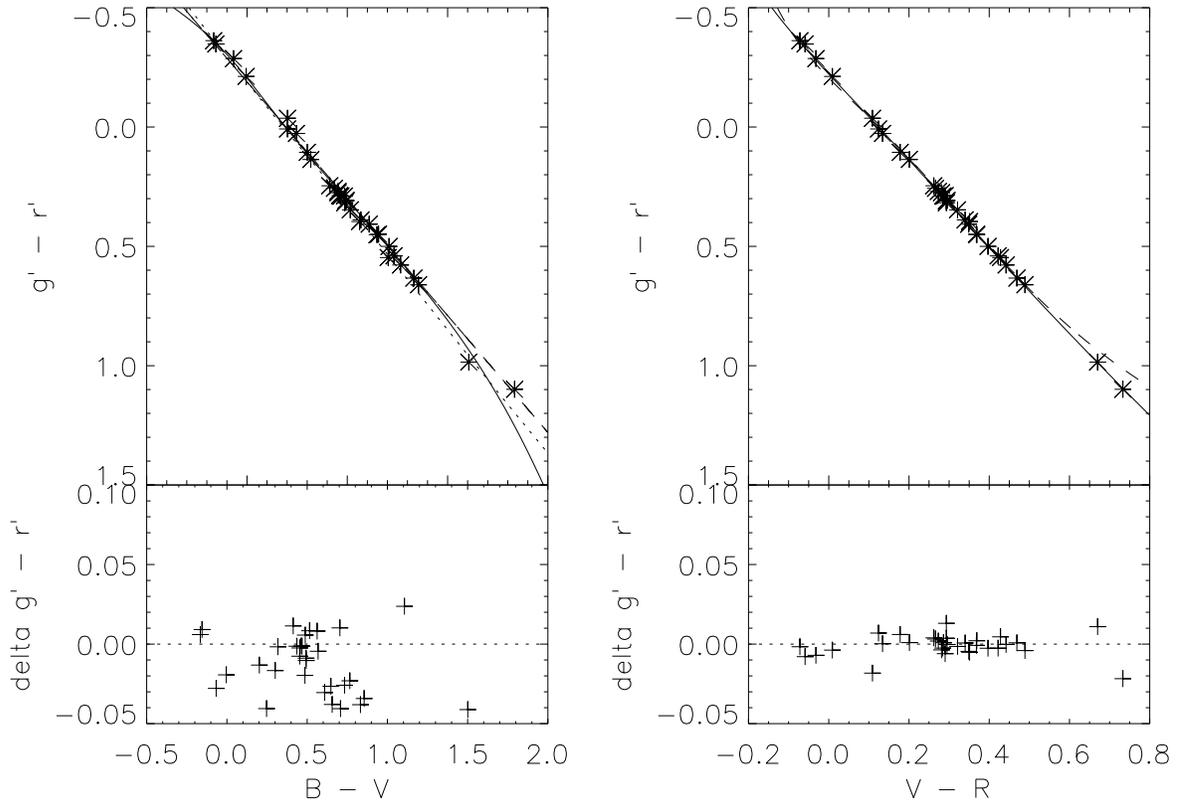}
\caption{The $g' - r'$ transformation equation fit shown here with a two 
dimensional cut of $g' - r'$ vs. \bv\ (left) and $V - R_C$ (right without
\citet{Karaali05} fit since there is no $V - R_C$ dependence in their
fits).  The lines 
in the top panels are the same as in Figure~\ref{trans_ub}.  The bottom panels
give the residuals for this paper's (right) and \citet{Karaali05} (left).  
\label{trans_gr}}
\end{figure}

Figure~\ref{trans_gr} shows the ZAMS fit of $g' - r'$ vs. $B - V$ 
and $g' - r'$ vs. $V - R_C$ with the residuals of \citet{Karaali05} and this paper.  Notice that \citet{Karaali05} residuals in the  $g' - r'$ vs. $V - R_C$ color-color plane are not available because \citet{Karaali05} did not use the $V - R_C$ color in their transformation equations.
The $g' - r'$ relation now becomes:

\begin{eqnarray}
g' - r' & = & 0.278 (\pm 0.016) (B - V) + 1.321 (\pm 0.030) (V - R_C) - 0.219\ .
\end{eqnarray}

\placefigure{trans_gr}

\subsection{The $r' - i'$ and $r' - z'$ Equations}

\begin{figure}
\includegraphics[scale=1.0]{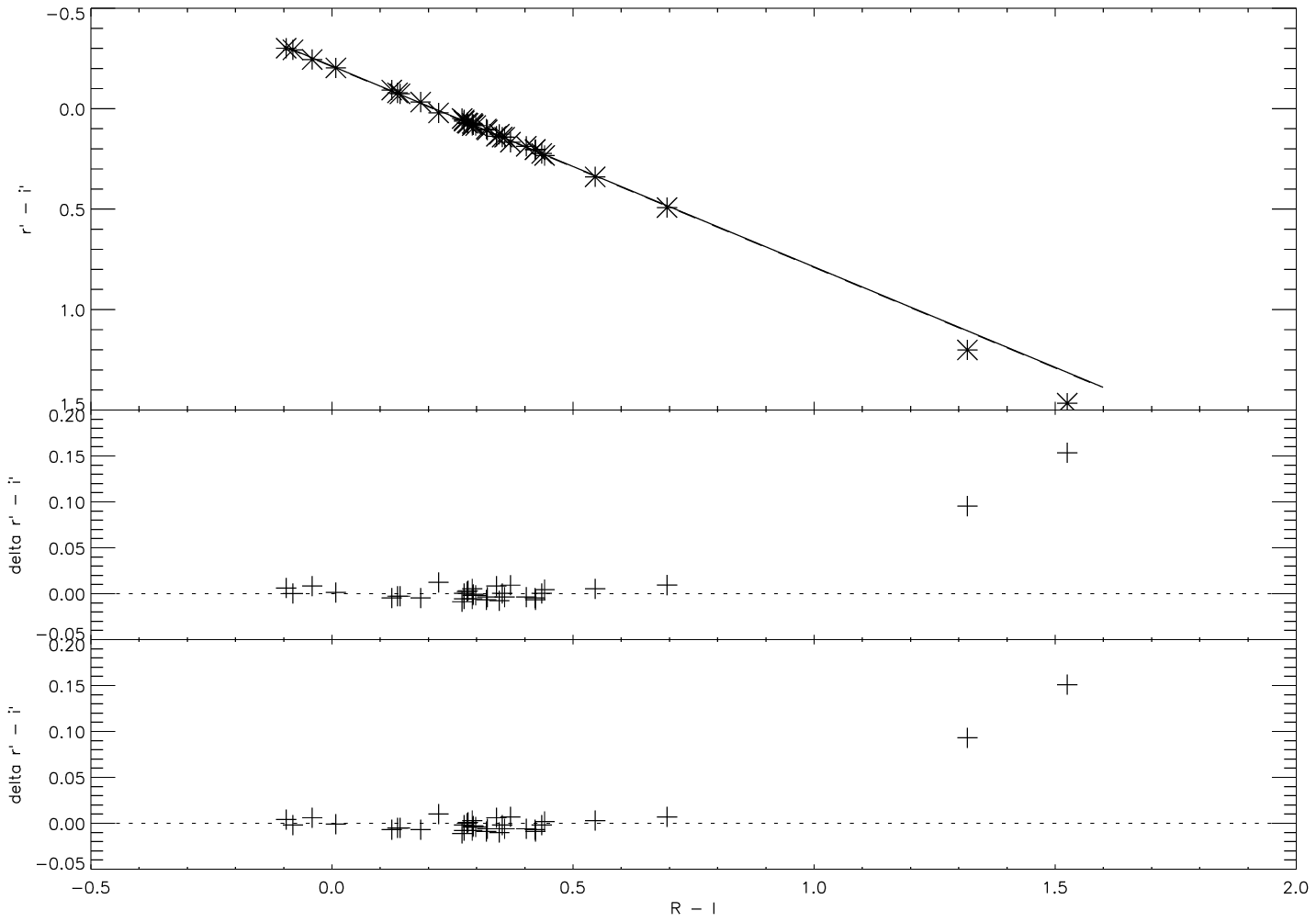}
\caption{The $r' - i'$ transformation equation fit with residuals using the 
\citet{Smith02} fit (middle) and this paper's fit (bottom).  The lines 
in the top panels are the same as in Figure~\ref{trans_ub}. The two extreme
red points were not included within the fit due to insufficient data for
stars with $B - V$ $>$ 1.3.
\label{trans_ri}}
\end{figure}

\begin{figure}
\includegraphics[scale=1.0]{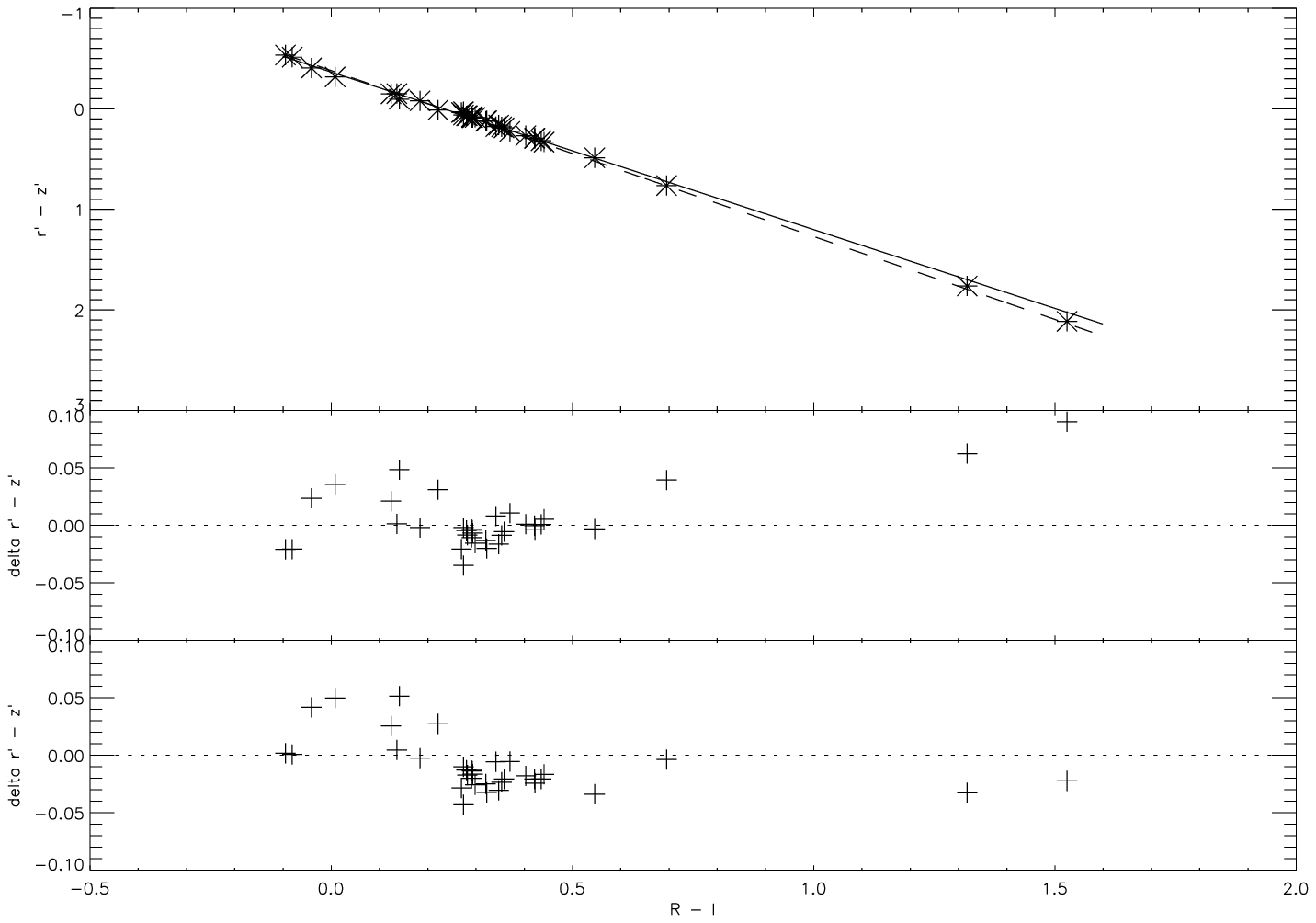}
\caption{The $r' - z'$ transformation equation fit with residuals using the 
\citet{Smith02} fit (middle) and this paper's fit (bottom).  The lines 
in the top panels are the same as in Figure~\ref{trans_ub}.  The two extreme
red points were not included within the fit due to insufficient data for
stars with $B - V$ $>$ 1.3.
\label{trans_rz}}
\end{figure}

Both the $r' - i'$ and $r' - z'$ relations are relatively simple 
compared to the previous two equations since the $r'$ and $R_C$ filters have
roughly the same effective wavelength and bandpass, and the $I_C$ filter is
mapped only by the $i'$ and $z'$ filters. For this reason, both of the 
SDSS colors 
will be dependent upon $R_C - I_C$ only.  We performed a least squares 
polynomial
fit to the data using the same in the $g' - r'$ data due to the lack of data for
TiO bands in M stars.
Figure~\ref{trans_ri} and Figure~\ref{trans_rz} show the $r' - i'$ vs. 
$R_C - I_C$
and $r' - z'$ vs. $R_C - I_C$ final fits with their respective residuals.  
Notice that the data
follow a well defined linear relation.  The equations for $r' - i'$ and 
$r' - z'$ are simply:

\begin{eqnarray}
r' - i' & = & 1.000 (\pm 0.006) (R_C - I_C) - 0.212\ , \\
r' - z' & = & 1.567 (\pm 0.020) (R_C - I_C) - 0.365\ .
\end{eqnarray}

\placefigure{trans_ri}
\placefigure{trans_rz}

\subsection{The $g' - V$ Equation}

\begin{figure}
\includegraphics[scale=1.0]{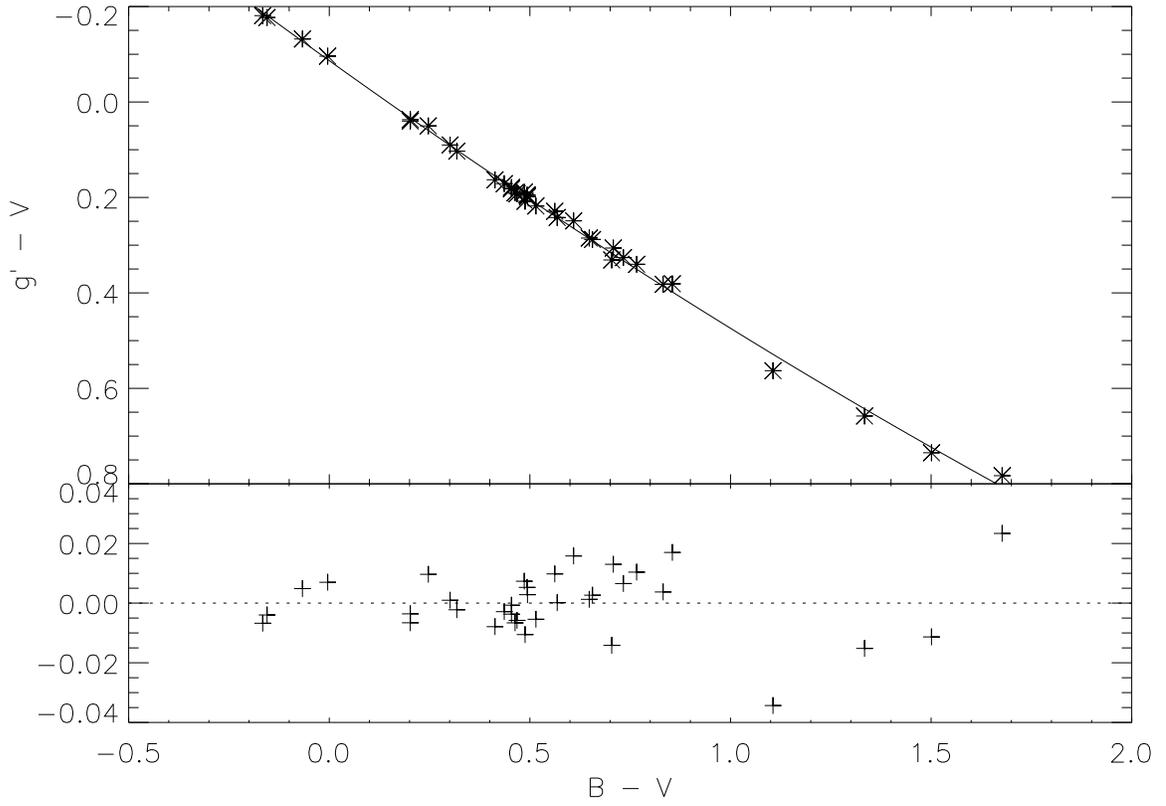}
\caption{The $g' - V$ transformation equation fit with residuals from this
paper's fit.   This was created for convenience of the user to acquire 
individual magnitudes in the $u'g'r'i'z'$ filter system.
\label{trans_gv}}
\end{figure}

The $g' -V$ color relation is derived mostly for convience to the user.  
In order to
obtain individual magnitudes in each of the $u'g'r'i'z'$ filters, we need
a logical color-color relation that will give the user the ability to
obtain one of the filter magnitudes.  Since the Johnson $V$ is often cited 
as the base magnitude for most published values,
the $g'$ filter is the most logical choice to
obtain $V$ magnitudes.  Since the $g'$ filter bandpass covers the $B$ and $V$ 
filter
bandpasses, the relation will be derived using a \bv color.  We did
a least squares polynomial fit to the data using all the available data in 
Table~\ref{stardata} excluding Ru-152.  Figure~\ref{trans_gv} illustrates the results 
of this fit with the residuals.  The equation for this relation is:

\begin{eqnarray}
g' - V & = & -0.042 (\pm 0.007) (B - V)^2 + 0.602 (\pm 0.011) (B - V) - 0.087\ .
\end{eqnarray}

\placefigure{trans_gv}

\section{Conclusions}
Comparing the residuals and the actual plots for each color transformation demonstrates that
our new transformation equations give better results for MS stars than \citet{Karaali05}
and \citet{Smith02}.
Additional work remain to fully develop the complete set of
transformation equations for all luminosity classes.  First, more 
data is needed for extremely red MS stars ($B - V$ $>$ 1.3).  Also, an
investigation of the effects of metallicity on the transformations
for each class of star must be undertaken.  This means that other 
isochrone models of $u'g'r'i'z'$ interest, with existing $UBVR_CI_C$ data 
(such as Y$^2$; \citep{Yi01}), 
must be identified and transformed.  This is a complicated and time
consuming prospect, but with the advent of high-precision photometry, 
it must be completed in order to deliver the accuracies needed when 
describing stellar systems.  

We plan to pursue this endeavor in future work. A beginning step is the
development of a full metallicity and luminosity photometric study in
the $u'g'r'i'z'$ filter system.  This study has all ready begun using the
0.6m telescope at Red Buttes Observatory in Laramie, Wyoming 
to support the extension of the Sloan Digital Sky Survey.

\acknowledgments
CTR, JAS, and DLT acknowledge support from National Science Foundation 
Grant No. 0098401.  JAS acknowledges support from the Los Alamos 
National Laboratory LDRD program 20030486DR.  RC, CTR, and MJP acknowledge
support from EPSCoR grant No. NCC5-578.

\newpage


\clearpage

\clearpage

\clearpage

\clearpage

\clearpage

\clearpage

\clearpage

\clearpage

\clearpage

\begin{deluxetable}{lrrrrrrrrr}
\tabletypesize{\footnotesize}
\tablecaption{Standard Star Data Set\label{stardata}}
\tablehead{
\colhead{Name} &
\colhead{$V$} &
\colhead{$B - V$} &
\colhead{$U - B$} &
\colhead{$V - R$} &
\colhead{$R - I$} &
\colhead{$u' - g'$} &
\colhead{$g' - r'$} &
\colhead{$r' - i'$} &
\colhead{$i' - z'$}
}
\startdata
92-342 & 11.613 & 0.436 & -0.042 & 0.266 & 0.270 & 1.069 & 0.257 & 0.049 & -0.012 \\
92-502 & 11.812 & 0.486 & -0.095 & 0.284 & 0.292 & 1.031 & 0.289 & 0.078 & 0.010 \\   
92-282 & 12.969 & 0.318 & -0.038 & 0.201 & 0.221 & 1.000 & 0.136 & 0.021 & -0.009 \\  
92-288 & 11.630 & 0.855 & 0.472 & 0.489 & 0.441 & 1.768 & 0.661 & 0.233 & 0.098 \\          
93-317 & 11.546 & 0.488 & -0.055 & 0.293 & 0.298 & 1.068 & 0.317 & 0.084 & 0.002 \\          
93-333 & 12.011 & 0.832 & 0.436 & 0.469 & 0.422 & 1.760 & 0.633 & 0.203 & 0.089 \\ 
94-242 & 11.728 & 0.301 & 0.107 & 0.178 & 0.184 & 1.157 & 0.106 & -0.033 & -0.046 \\          
95-96  & 10.010 & 0.147 & 0.072 & 0.079 & 0.095 & 1.142 & -0.070 & -0.123 & -0.051 \\ 
95-218 & 12.095 & 0.708 & 0.208 & 0.397 & 0.370 & 1.442 & 0.500 & 0.167 & 0.058 \\ 
95-236 & 11.491 & 0.736 & 0.162 & 0.420 & 0.411 & 1.414 & 0.539 & 0.196 & 0.084 \\ 
96-36  & 10.591 & 0.247 & 0.118 & 0.134 & 0.136 & 1.196 & 0.027 & -0.079 & -0.072 \\         
96-737 & 11.716 & 1.334 & 1.160 & 0.733 & 0.695 & 2.770 & 1.099 & 0.492 & 0.271 \\          
96-83  & 11.719 & 0.179 & 0.202 & 0.093 & 0.097 & 1.237 & -0.054 & -0.110 & -0.048 \\        
97-249 & 11.733 & 0.648 & 0.100 & 0.369 & 0.353 & 1.317 & 0.451 & 0.141 & 0.038 \\        
97-351 & 9.781 & 0.202 & 0.096 & 0.124 & 0.141 & 1.130 & 0.008 & -0.074 & -0.022 \\     
98-978 & 10.572 & 0.609 & 0.094 & 0.349 & 0.322 & 1.277 & 0.407 & 0.106 & 0.013 \\    
98-185 & 10.536 & 0.202 & 0.113 & 0.109 & 0.124 & 1.153 & -0.037 & -0.093 & -0.057 \\    
98-653 & 9.539 & -0.004 & -0.099 & 0.009 & 0.008 & 0.843 & -0.212 & -0.203 & -0.114 \\   
98-685 & 11.954 & 0.463 & 0.096 & 0.290 & 0.280 & 1.218 & 0.287 & 0.070 & -0.001 \\    
Ru-152 & 13.014 & -0.190 & -1.073 & -0.057 & -0.087 & -0.263 & -0.355 & -0.289 & -0.252 \\      
99-438 & 9.398 & -0.155 & -0.725 & -0.059 & -0.081 & 0.136 & -0.348 & -0.293 & -0.220 \\    
99-447 & 9.417 & -0.067 & -0.225 & -0.032 & -0.041 & 0.734 & -0.287 & -0.245 & -0.161 \\     
100-241 & 10.139 & 0.157 & 0.101 & 0.078 & 0.085 & 1.165 & -0.068 & -0.128 & -0.097 \\      
100-280 & 11.799 & 0.494 & -0.002 & 0.295 & 0.291 & 1.143 & 0.308 & 0.084 & 0.003 \\      
BD-12:2918 & 10.067 & 1.501 & 1.166 & 1.067 & 1.318 & 2.817 & 1.326 & 1.201 & 0.561 \\        
101-316 & 11.552 & 0.493 & 0.032 & 0.293 & 0.291 & 1.152 & 0.309 & 0.073 & 0.007 \\    
101-207 & 12.419 & 0.515 & -0.078 & 0.321 & 0.320 & 1.085 & 0.347 & 0.101 & 0.022 \\   
103-626 & 11.836 & 0.413 & -0.057 & 0.262 & 0.274 & 1.056 & 0.246 & 0.056 & -0.027 \\   
104-598 & 11.479 & 1.106 & 1.050 & 0.670 & 0.546 & 2.481 & 0.985 & 0.339 & 0.148 \\   
DM+2\_2711 & 10.367 & -0.166 & -0.697 & -0.072 & -0.095 & 0.166 & -0.362 & -0.301 & -0.234 \\    
107-1006 & 11.712 & 0.766 & 0.279 & 0.442 & 0.421 & 1.549 & 0.578 & 0.204 & 0.090 \\   
107-351 & 12.342 & 0.562 & -0.005 & 0.351 & 0.358 & 1.187 & 0.396 & 0.142 & 0.048 \\  
108-551 & 10.703 & 0.179 & 0.178 & 0.099 & 0.110 & 1.256 & -0.032 & -0.104 & -0.051 \\ 
Wolf629 & 11.759 & 1.677 & 1.256 & 1.185 & 1.525 & 3.013 & 1.413 & 1.466 & 0.648 \\  
109-381 & 11.730 & 0.704 & 0.225 & 0.428 & 0.435 & 1.477 & 0.547 & 0.223 & 0.094 \\  
112-223 & 11.424 & 0.454 & 0.010 & 0.273 & 0.274 & 1.145 & 0.270 & 0.062 & 0.000 \\  
112-805 & 12.086 & 0.152 & 0.150 & 0.063 & 0.075 & 1.183 & -0.087 & -0.135 & -0.090 \\   
113-339 & 12.250 & 0.568 & -0.034 & 0.340 & 0.347 & 1.149 & 0.389 & 0.127 & 0.035 \\   
113-466 & 10.004 & 0.454 & -0.001 & 0.281 & 0.282 & 1.125 & 0.275 & 0.073 & -0.005 \\   
114-531 & 12.094 & 0.733 & 0.186 & 0.422 & 0.403 & 1.419 & 0.540 & 0.187 & 0.080 \\   
114-654 & 11.833 & 0.656 & 0.178 & 0.368 & 0.341 & 1.398 & 0.449 & 0.137 & 0.040 \\   
114-750 & 11.916 & -0.041 & -0.354 & 0.027 & -0.015 & 0.548 & -0.212 & -0.230 & -0.163 \\    
115-420 & 11.161 & 0.468 & -0.027 & 0.286 & 0.293 & 1.091 & 0.290 & 0.080 & 0.007 \\   
115-516 & 10.434 & 1.028 & 0.759 & 0.563 & 0.534 & 2.167 & 0.807 & 0.317 & 0.172 \\   
\enddata
\end{deluxetable}

\clearpage

\begin{deluxetable}{cc}
\tablewidth{1.75in}
\tabletypesize{\footnotesize}
\tablecaption{Reduced $\chi^2$ \label{chisqr}}
\tablehead{
\colhead{Color} &
\colhead{Reduced $\chi^2$}
}
\startdata
$u' - g'$ & 0.0003\\
$g' - r'$ & 0.001\\
$r' - i'$ & 0.001\\
$r' - z'$ & 0.011\\
\enddata
\end{deluxetable}
\end{document}